\documentclass[11pt,twoside]{article}
\usepackage{asp2010}

\pdfoutput=1

\resetcounters

\begin{document}

\title{Integrated Properties of AGB Stars in Resolved and Unresolved Stellar 
Populations : Simple Stellar Populations and Star Clusters}
\author{Ariane~Lan\c{c}on$^1$
\affil{$^1$Observatoire Astronomique de Strasbourg, 
 Universit\'e de Strasbourg, CNRS, UMR 7550,
 11 rue de l'Universit\'e, F-67000 Strasbourg, France}}

\begin{abstract}
The evolution of AGB stars is notoriously complex. The confrontation of 
AGB population models with observed stellar populations is a useful 
alternative to the detailed study of individual stars in efforts to converge 
towards a reliable evolution theory. I review here the impact of 
studies of star clusters on AGB models and AGB population synthesis, 
deliberately leaving out any more complex stellar populations. Over the 
last 10 years, despite much effort, the absolute uncertainties in the 
predictions of the light emitted by intermediate age populations have 
not been reduced to a satisfactory level. Observational sample definitions, 
as well as the combination of the natural variance in AGB properties with 
small number statistics, are largely responsible for this situation. 
There is hope that the constraints may soon become strong enough, 
thanks to large unbiased surveys of star clusters, 
resolved colour-magnitude diagrams, and new analysis methods 
that can account for the stochastic nature of AGB populations in clusters.
\end{abstract}

\section{The Upper AGB in Population Synthesis Models}

With the advent of near-IR astronomy on one hand and the efforts devoted to
the modelling of the chemical evolution of galaxies on the other, it has 
rapidly become obvious that AGB stars must be included very carefully
in all varieties of population synthesis models 
\citep{Iben_Truran_1978, Renzini_Buzzoni_1986, Charlot_Bruzual_1991}.  
Meetings such as the IAU Symposium held in Montpellier in 1998 or the 
first edition of the Vienna meetings in 2006 (GALAGB1 hereafter) 
highlighted the reasons for such an effort and the difficulties to overcome. 
Today the whole extragalactic community has been warned that stars on the 
upper AGB (to which we will refer as the TP-AGB hereafter because 
it is usually associated with thermal pulses) is responsible for more than
50\,\% of the near-IR light emitted by a stellar population at intermediate
ages (a few 100\,Myr to one or two Gyr). It may well reach 80\,\% at a
range of ages and metallicities \citep{Maraston_2005,Marigo_etal_2010_IAUS}. 
AGB stars determine the near-IR
mass-to-light ratio of these populations, with critical effects on the
estimates of galaxy masses when the universe had an age of order 
$10^9$ years \citep{Maraston_Daddi_etal_2006}. 

For this second edition of the Vienna meetings, I was asked to 
review the current status of population synthesis models for 
Simple Stellar Populations (SSPs) at ages at which AGB stars are most relevant.
This paper therefore excludes any discussion of composite populations. 
And it focuses on star clusters as approximate SSPs in the real world although
there is growing evidence that the fraction of clusters with
not-so-simple populations is much larger than one dared to hope a few years 
ago \citep{Gratton_Sneden_Carretta_2004,Milone_etal_2009}. 
To narrow down the field even more, only models and 
observations at optical and near-IR wavelengths will be considered, 
despite the importance of AGB envelopes at the mid-IR wavelengths 
\citep{Bressan_etal_1998,Vega_Bressan_etal_2010} 
and the debated role of post-AGB stars in the ultraviolet
\citep[e.g.][]{BrownT_etal_2008}.

\subsection{From stellar evolution models to stellar populations}

Most of the population synthesis codes used in the extragactic community
include the AGB through so-called synthetic modelling, i.e. 
a set of analytic scaling relations that reproduce the main 
behaviours of detailed interior and evolution models, because computation
times still limit the size of the grids that can be produced with
the latter. A list of existing synthetic evolution codes has been compiled recently by 
\citet{Marigo_etal_2010_IAUS}.
A minimal set of ingredients of synthetic models includes a core mass -
luminosity relation, a link between the core growth rate and luminosity,
the rate of evolution for the total mass and therefore some 
prescription for mass loss (which frequently implies estimating 
a radius or effective temperature), an approximate description of the pulse
shapes and the interpulse durations, a description of the 
impact of 3rd dredge-up and the formation of carbon stars. 
Both core burning and envelope burning must be accounted for.
The most frequently used  set of basic relations still is the one by
Wagenhuber \& Groenewegen (1998), which is based on the
original tracks of Wagenhuber \& Weiss (1994) and Wagenhuber (1996).
Another starting point is \citet{Hurley_Pols_Tout_2000}.
Modern implementations all include a dependence on metallicity. 
Some consider abundance ratios, in particular the large effects of
the C/O ratio on envelope opacities, which sets the stellar radii and
therefore influences mass loss rates and stellar lifetimes. Many of 
the prescriptions listed above have modern versions with several 
parameters each, that must be considered essentially free in view of the
theoretical uncertainties. This is what is meant in population synthesis 
jargon when it is said that models need to be calibrated against observations.
With more parameters being added to the models, this task 
is not actually becoming easier as time passes: the number of
independent empirical constraints has to grow equally fast. This point 
is discussed further in Sect.\,\ref{ALancon_Calibrating.sec}.

A few alternatives to synthetic evolution are being used today. One approach
is directly based on the ``fuel consumption theorem". This approach 
bypasses the calculation of the time-dependent location of evolving stars
in the HR-diagram, and focuses on the total energy radiated by 
each (spectral) type of post-main-sequence star for a given main-sequence turn-off
\citep[as originally done by][]{Tinsley_Gunn_1976}.
The number of free model parameters is kept small and the main scaling relations are 
taken from observations of star cluster samples whenever possible 
\citep{Maraston_2005}. The opposite approach,
meant to force agreement with nature and highlight shortcomings of
current synthetic models, with no claim to ensure physical consistency,
is to add even more free parameters to existing population synthesis models and 
vary them radically \citep{Conroy_Gunn_2010}. Finally, at least one group
is attempting to use modern CPU grids to produce enough 
complete stellar interior and evolution models to avoid the need 
for analytic prescriptions for interpolation and extrapolation
(Puzia et al., in preparation). The ressources needed 
to work with this approach are costly even for today's standards, and of course 
the fundamental uncertainties related among others to  the choice of a 
criterion defining convective boundaries, of a set of nuclear reaction rates, 
and of a mass loss prescription are not removed.

\begin{figure}
\centerline{\includegraphics[clip=,width=0.5\textwidth]{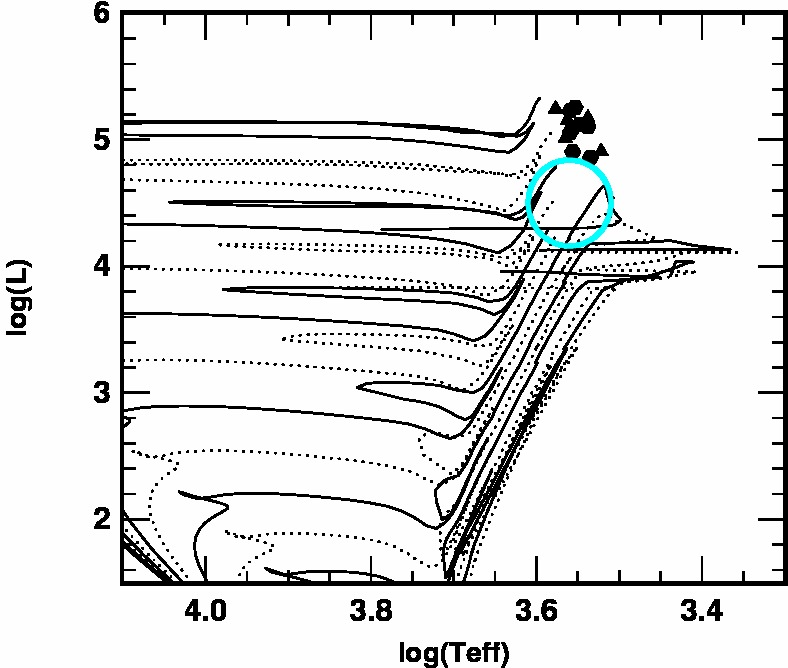}
}
\caption[]{The super-AGB gap. The isochrones shown for Z=0.008 and ages between 10\,Myr
and 4\,Gyr are from \citet{Marigo_etal_2008};
the thermal pulses of super-AGB stars are not included. Symbols represent the LMC/SMC red
supergiant sample of Levesque et al. (2007). The light coloured ellipse highlights the 
typical ``zone of avoidance" of empirical samples of either red supergiants or AGB stars.}
\label{ALancon_superAGBgap.fig}
\end{figure}

It takes some time for the latest findings derived from specialised 
stellar evolution calculations to spread out into the realm of 
stellar population studies. One example of such a delay is seen currently
in the absence of super-AGB stars in widely used population synthesis codes.
Super-AGB stars form the boundary between traditional low or intermediate
mass stars that will form carbon-oxygen white dwarfs, and massive stars
that will explode as supernovae after burning not only carbon but also 
its products. They are predicted to manage to burn only 
carbon before either forming neon-oxygen white dwarfs or exploding
as electron capture supernovae \citep[e.g.][]{Ritossa_etal_1996,Poelarends_etal_2008,
Siess_TP-SAGB_2010}. 
They undergo thermal pulses before they die, and become 
extremely luminous ($\sim 10^5$\,L$_{\odot}$), but 
in many (all?) current population synthesis codes the stars simply disappear when
the available evolutionary tracks end, i.e. before the thermal pulses.
A symptom of the disarray is that empirical samples of red supergiants
on one hand, bright AGB stars on the other, tend to carefully avoid
the region of confusion, and that it therefore remains difficult to find
samples of candidate super-AGB stars (Fig.\,\ref{ALancon_superAGBgap.fig}).

\subsection{From stellar populations to emission properties}

The output of evolutionary calculations, combined with a stellar
initial mass function (IMF) and the assumption of a single stellar
population (all stars have equal age and initial composition), is 
a distribution of stars across evolutionary stages, which is generally
made available as a set of luminosities (or gravities), effective temperatures
and surface abundances. Predicting the emission of such a population
when TP-AGB stars contribute strongly remains a serious challenge.  

TP-AGB stars are cool, and model atmosphere calculations coupled with 
detailed radiative transfer codes in spherical symmetry have been
demonstrated to reproduce observations well for static M-type giants only 
above about 3500\,K. Even in that range of temperature the demonstration of 
the models' ability to match high resolution optical and near-IR spectra 
simultaneously is still absent due to lack of adequate reference observations. 
Projects such as the ongoing Xshooter spectral library could help here
(PI S. Trager, see Chen et al. in this volume). At cooler temperatures,
the models seize the main trends but do not match optical and near-IR data 
with the level of consistency required for common stellar population studies.
A main cause for trouble is that real cool giant and supergiant stars are 
all variable in one way or another, with irregular variations due
for instance to surface inhomogeneities related to surface convection, or
for luminous AGB stars long period variations associated with pulsation.
Models for the spectra of Mira-type pulsators on the TP-AGB are
in their youth \citep{Tej_etal_2003,Lebzelter_etal_2010}, 
and the continuing efforts of the few teams devoting their time 
to these calculations and their physical ingredients are recognized by
the population synthesis community as extremely important.  
As summarized by S. H\"ofner (this volume), current models remain 
highly simplified, and rather than on the output spectra they 
still focus on evolutionary aspects, in particular the formation of dust grains 
capable of driving the wind that leads to the ejection of the envelope; 
nevertheless, with new ideas on the nature of the dust particles 
that might play this role, late M-type TP-AGB stars receive fresh 
attention. Non-linear pulsation models, such as constructed
by \citet{Olivier_Wood_2005} for instance, are not as yet combined
with detailed atmosphere and radiative transfer models.

The principles of dust formation in TP-AGB stars with 
carbon-rich atmospheres had been set before GALAGB1.
Spectral models for carbon stars exist and are continuously
being upgraded (Nowotny, this volume). The grid of static models of \citet{Aringer_etal_2009}
only contains carbon stars with relatively blue optical to near-IR colours. 
New calculations with (piston-driven) pulation models, that produce circumstellar
dust, reproduce the range of carbon star colours observed in nature
(Eriksson et al., this volume). 

The alternative to theory is to work with empirical spectra of AGB stars.
The largest collection of empirical spectra directly useful for population
synthesis purposes was produced by \citet{Lancon_Wood_2000} and 
\citet{Lancon_Mouhcine_2002_avgspec}. It is included in the spectral synthesis
models of \citet{Mouhcine_Lancon_I_2002,Maraston_2005, Raimondo_2009, Percival_etal_2009}.
The Xshooter spectral library already contains observations of about a hundred
such stars.

It must be kept in mind that empirical spectra of long period variables (LPVs) 
don't carry a label indicating the fundamental parameters of the 
observed stars. Without a final theory for the time-dependent emission 
of LPVs, the assignment of an effective temperature
or a mass can not be very reliable. Let alone the assignment of the 
effective temperature of the ``static parent" of the star. But 
just that is what is needed to attach an empirical spectrum (time-averaged
over the pulsation cycle) to a particular point of a theoretical
stellar evolution track or isochrone. Tracks have enough trouble dealing with 
thermal pulses and never consider the {\em systematic} 
effect pulsation is expected to have on the average effective temperature
of a star \citep[e.g.][]{Hofmann_etal_1998}. Illustrations of the issues are provided in 
\citet{Lancon_Mouhcine_2002_avgspec}. The uncertainties associated with
the connection between a point on a theoretical track and a 
representative spectrum on the TP-AGB are {\em larger} than those associated 
with the value of the theoretical (non-pulsating) effective temperature. 

With thousands of observations of LPVs across optical and near-IR wavelengths,
rather than hundreds, it should become possible to define better  
correlations  between spectral properties and pulsation properties
(pulsation mode, period, amplitude) than available today.
Such a project would take a long
time, but it seems that obtaining the equivalent theoretical sequences
of intermediate resolution spectra will take even longer. 
Pulsation models can predict periods along evolutionary tracks, maybe also
amplitudes, which in the end could help selecting the right spectrum
for any given evolutionary stage. This idyllic path into the future 
darkens a bit if one remembers that stability against pulsation and 
pulsation properties are sensitive to the very outer stellar layers and to
poorly understood physics such as convection (see Wood, this volume).

At present, carbon stars seem somewhat easier to deal with than 
oxygen-rich LPVs. Indeed, the variance between the 
spectral features observed in carbon star samples is much smaller
than for their oxygen-rich counterparts. 
The question of S-stars has 
until now been avoided in the population synthesis community by
arguing that this phase of evolution is short compared to others,
which is the case if an S-star is defined as having a C/O 
surface abundance ratio very close to 1.
At optical wavelengths, S-stars differ from M-stars as soon as 
C/O $\simeq$ 0.5 (van Eck, this volume). 
More relevant to the emission of a stellar population as 
a whole is what happens around 1\,$\mu$m or beyond. It is probably
worth reconsidering possible effects.

\section{Calibrating SSP models against observations of 
intermediate age clusters}
\label{ALancon_Calibrating.sec}

The calibration of AGB population synthesis models against 
observations is the determination of a set of model assumptions 
and parameters that optimizes the correspondance with available data.
Typical sets of data that have been used (often jointly) in this 
process include the luminosity functions of AGB stars and of carbon 
stars in the Magellanic Clouds, the ratio of M stars to C stars
in various Local Group galaxies, the abundance patterns 
seen in planetary nebulae. The colours of
high redshift galaxies, which are not contaminated by any stars older
than a few Gyr, also provide direct constraints on the energy released
by stars  on the AGB. Finally, resolved HR diagrams of 
nearby galaxy field stars will increasingly contribute to providing constraints 
on AGB models.

Star clusters play a fundamental role in the calibration stellar population
models. The ages at which AGB stars contribute to the light budget
are spread over a range wide enough that one can expect useful 
constraints  even with internal age spreads of order $10^8$\,yr
\citep{Milone_etal_2009}.
The effects of internal abundance spreads, rotation and binarity
on the TP-AGB model calibration process remain to be evaluated.

The cluster systems of the two Magellanic Clouds are traditional calibration
samples \citep{Ferraro_etal_1995,Maraston_1998,Mouhcine_Lancon_I_2002}. 
In the Clouds, \citet{Bica_etal_2008} identified 
thousands of clusters and associations, 
and hundreds of these are thought to have intermediate ages 
\citep{Chiosi_etal_2006,Glatt_Grebel_Koch_2010,Chandar_etal_2010_LMC1}. 
However, the samples used for the calibration of currently used
population synthesis tools include only a few tens of these
objects.  In addition, the selection criteria defining the samples 
are usually described too superficially to allow a proper study of biases. 

A major, very serious caveat of the Magellanic Cloud clusters is that 
only very few high mass clusters are found among them. 
Mass distributions of star clusters
in galaxies tend to fall steeply with increasing mass, therefore it takes
very large ensembles of clusters to find many massive ones. 
In the Milky Way, the number of massive intermediate age open clusters
is even lower \citep{Popescu_Hanson_2010_MasscleanCols}. Comparisons with other intermediate
age cluster samples remain relatively rare \citep[e.g.][among others]{Vazquez_Leitherer_2005,
Mouhcine_etal_2002_W3}. 

What clusters can be 
considered ``massive enough" for the particular task of calibrating 
AGB models? Clusters are mostly used to test the age and metallicity 
dependence of model colours. $V-K$ is directly sensitive to the
total contribution of cool stars (they contribute little in V and
very strongly in K), while near-IR colours such as J-K help constraining
the nature of the cool stars (O-rich or C-rich, effective temperature).
A single cluster is massive enough if its colours can be considered 
representative of a population of a given age and metallicity. 
TP-AGB stars are both luminous and intrinsically rare. With 
standard stellar initial mass functions, there will be 1 TP-AGB star 
for some $10^4$ stars in total, and that one will produce 70\,\% of the
K band light. Poisson statistics tells us that it will take about 
100 TP-AGB stars, i.e. (model-dependent) total masses around 
$5\,10^5$\,M$_{\odot}$, to restrict natural dispersion in TP-AGB 
numbers (between clusters of the same age, metallicity and total mass) 
to less than 10\,\% of the average 
\citep{Lancon_Mouhcine_2000_stoch}. 
This is also the condition for one cluster to display a $V-K$ colour 
within 10\,\% of the mean.  
At lower masses, only a handful of TP-AGB stars produce most of the
clusters' near-IR light, and both the exact number and the current 
spectral type of these will have strong direct effects on the clusters' 
spectra and colours.
Awareness of the so-called ``stochastic fluctuation problem" and the 
non-gaussian colour distributions that result from it, dates back
to the 1970s  
\citep{Barbaro_Bertelli_1977,Girardi_Bica_1993,Santos_Frogel_1997}. 

A proposed solution to this problem has been to sum individual 
observed clusters, thus constructing ``superclusters" with total masses 
of order $10^6$\,M$_{\odot}$. \citet{Gonzalez-Lopezlira_etal_2005} and
\citet{Pessev_etal_2008} have invested considerable
efforts in producing supercluster data that include near-IR fluxes. The trend
is that these supercluster don't display colours nearly as red as 
those that were thought representative based on individual clusters: 
$V-K$ peaks around 2.5 for Magellanic Cloud superclusters, instead of 3.5 in 
the sample of individual clusters most frequently used previously 
\citep{Persson_etal_1983}.
Difficulties in matching optical and near-IR apertures for 
the integrated photometry of star clusters on a crowded 
background explain parts of the discrepancy \citep{Pessev_etal_2008}.
Several of the current population synthesis models do reach 
$V-K=3.4$ at Magellanic Cloud metallicities 
\citep{Maraston_2005,Marigo_etal_2008}.
Also, the reddest colour occurs shortly after $10^8$\,yr
of evolution in those models, while the ages assigned to the 
reddest empirical superclusters (based on optical colours, spectra
or resolved colour-magnitude diagrams) are of about 1\,Gyr. 
Thus, it seems that the calibration process of the TP-AGB population
synthesis models needs to be reconsidered 
\citep[for further discussion, see][]{Gonzalez-Lopezlira_etal_2010,
Pessev_etal_2008,Conroy_Gunn_2010}. As we will now argue, work on 
the cluster data will be as important as work on the models.

\begin{figure}
\includegraphics[width=0.49\textwidth]{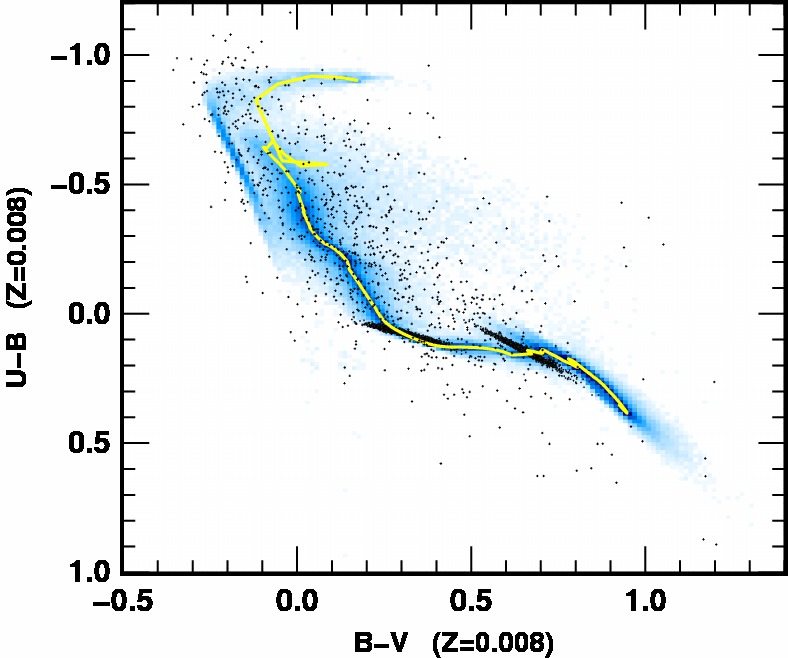}
\includegraphics[width=0.49\textwidth]{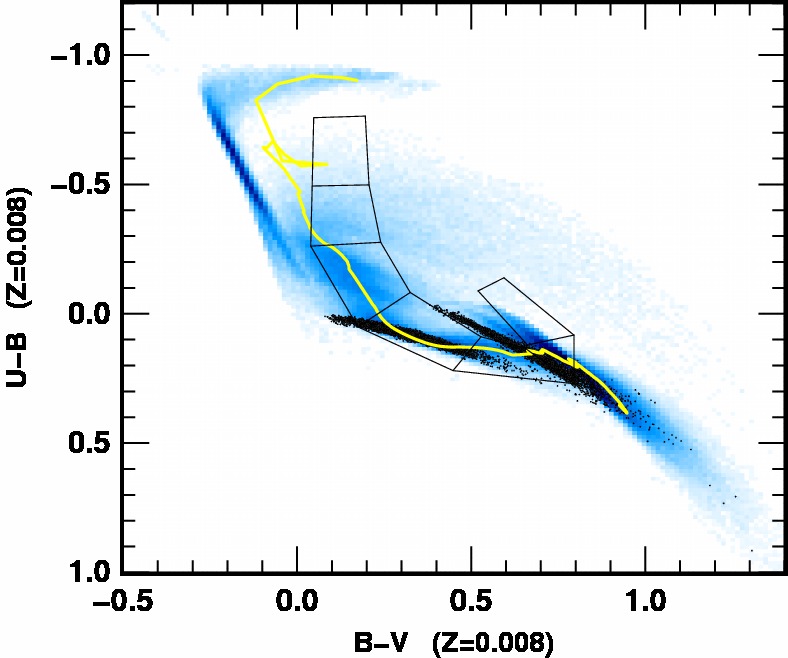}
\includegraphics[clip=,width=0.49\textwidth]{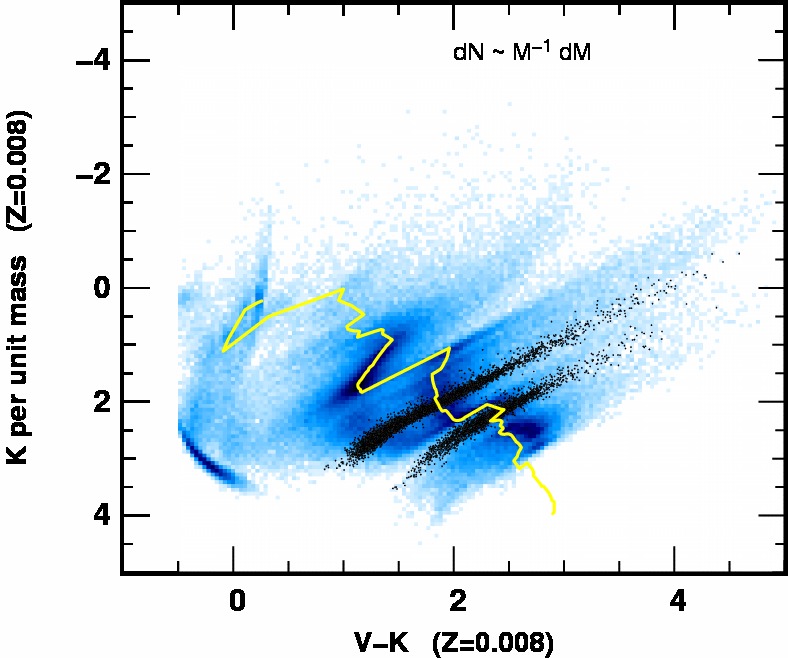}
\includegraphics[clip=,width=0.49\textwidth]{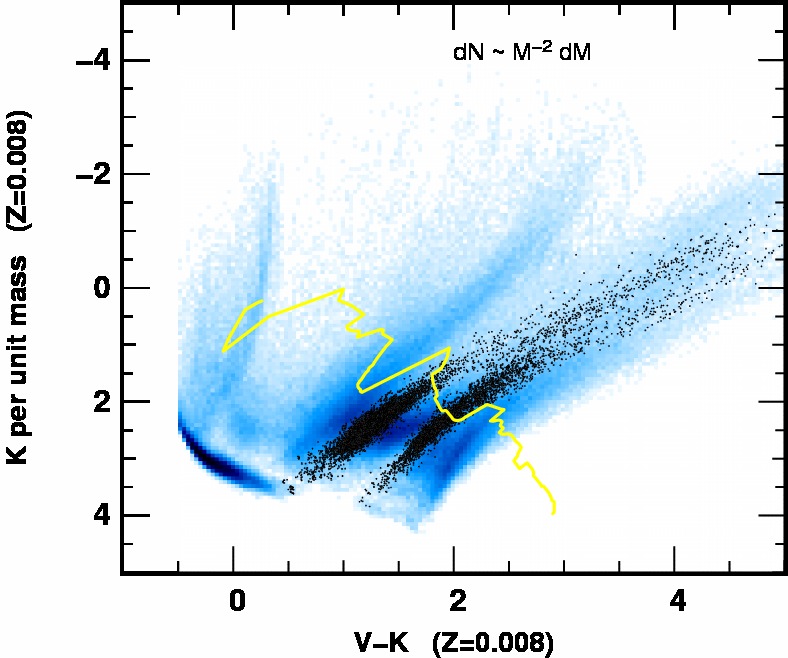}
\includegraphics[clip=,width=0.49\textwidth]{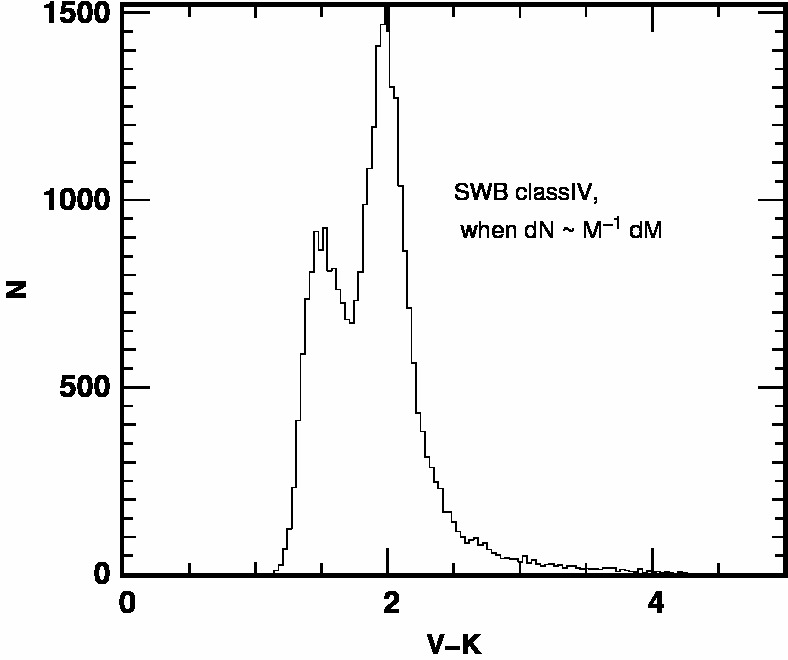}
\rule[0pt]{3pt}{0pt}
\includegraphics[clip=,width=0.49\textwidth]{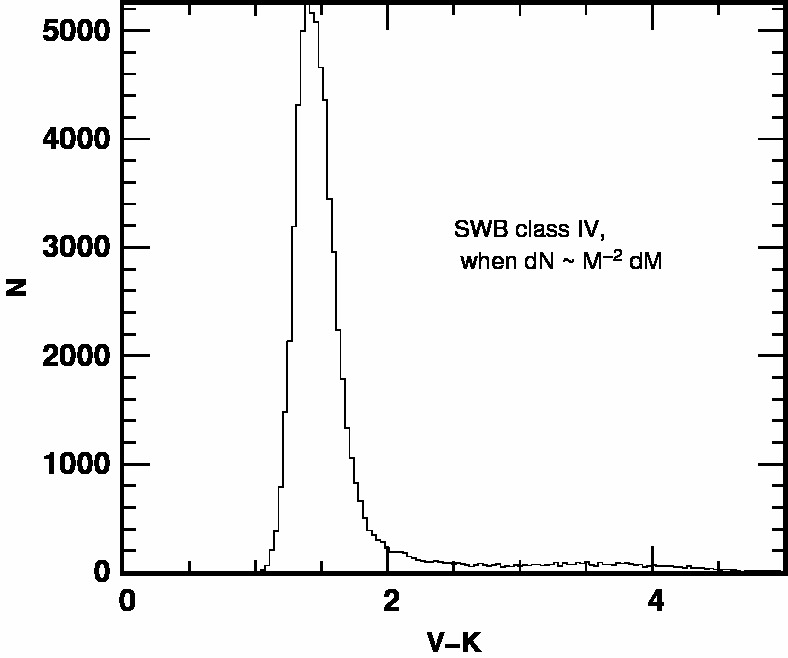}
\caption[]{{\em Top:}\ The integrated colours of clusters are 
relatively well-behaved at the optical wavelengths 
frequently used for approximate age-dating. Displayed
are the loci of models at Z=0.008 for two cluster 
mass distributions ($M^{-1}$ on the left, $M^{-2}$ on the right). 
Overlaid in the left panel are LMC clusters of \citet{Hunter_etal_2003}.
The boxes in the right panel delineate the SWB-classes I to VII  of
\citet{Frenk_Fall_1982}, corrected approximately to zero reddening). 
Clusters with ages of 500\,Myr and 1.5\,Gyr
are highlighted in black. Solid line\,: locus of very massive clusters.
{\em Middle:}\ Near-IR fluxes are strongly affected by rare luminous red
stars such as TP-AGB stars. Clusters with given UBV properties will
display a wide range of $V-K$ colours and $K$-band mass-to-light ratios.
{\em Bottom:}\ $V-K$ colour distribution predicted for collections of 
star clusters of SWB class IV (i.e. intermediate age clusters).
}
\label{AL_StochasticClusters.fig}
\end{figure}

The ``supercluster" procedure is a valid method of reducing errors on the
estimate of the mean colours {\em as long as no biases are introduced}
when selecting and age-dating the cluster samples used as input.
In current samples, selection biases have not been described well enough.
Mostly, the choice has been
driven by technical considerations : compactness, a reasonably uniform
and not too crowded stellar background, no obvious dust lanes across
the cluster image, etc. Characterizing selection biases
and attempting to construct unbiased samples will take considerably more work,
as it requires extensive artificial cluster experiments, or at least
a study of the effects of magnitude limits on clusters with finite
numbers of stars. Large collections of discrete synthetic clusters
are becoming available for such work 
\citep{Deveikis_etal_2008,CervinoVallsGabaud_2009,Popescu_Hanson_2010_MasscleanCols,Fouesneau_Lancon_2010}.

Discrete synthetic clusters help illustrating why the risk of biases is strong.
Figure\,\ref{AL_StochasticClusters.fig} shows the predicted loci of clusters 
in colour-colour space at Z=0.008, the metallicity
of intermediate age LMC-clusters.
When most of the clusters are small, only very few are expected to
lie along the line that marks average properties: meeting the average
with a small cluster would require a non-integer number of red stars.
In view of the natural dispersion
in the colours at a given age, the UBV plane allows us to
group observed intermediate age clusters into no more than four or five bins of 
intermediate ages.  Note that extinction moves clusters along the age sequence 
and blurs the picture \citep[see also][]{Girardi_etal_1995}, and that 
uncertain stellar model ingredients such as overshooting affect 
main sequence lifetimes and the age-colour relation of isochrones.  

In the near-IR, colour distributions are highly asymetric. The models
used in Figure\,\ref{AL_StochasticClusters.fig} have the assumptions
described in \citet{Fouesneau_Lancon_2010}. They include
a very simple TP-AGB based on \citet{Groenewegen_deJong_1993}\footnote{As in 
P\'egase.2, ftp://ftp.iap.fr/pub/from\_users/pegase/PEGASE.2/}. With
these models, the populous clusters have $V-K$ between 1.7 and 2.4. 
However, lower mass cluster $V-K$ colours are spread 
between 1.3 and 4.5, depending on the exact number and nature of TP-AGB stars 
they happen to contain. Many have no TP-AGB star and are very blue. It is
easy to imagine observational biases, in favour of clusters with either more 
or fewer TP-AGB stars than the average. Contamination by field stars 
adds uncertainties to the colours. 

\begin{figure}
\includegraphics[clip=,width=0.49\textwidth]{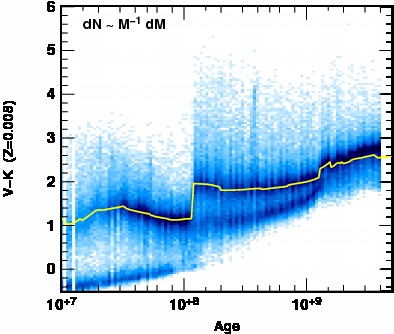}
\includegraphics[clip=,width=0.49\textwidth]{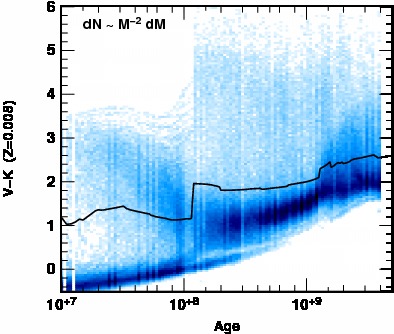}
\caption[]{The run of $V-K$ versus age for discrete star clusters
at Z=0.008, with two cluster mass distributions (solid line\,: most massive clusters). 
Synthetic clusters produced by M.  Fouesneau, as described in \citet{Fouesneau_Lancon_2010}.
}
\label{AL_VmK_age.fig}
\end{figure}

In the future, it will make sense to  combine the use
of ``superclusters" with the analysis of complete star cluster populations of
galaxies. The comparison of Figure\,\ref{AL_VmK_age.fig} with those
of other authors, for instance with Fig.\,8 of \citet{Marigo_etal_2008}
or Fig.\,2 of \citet{Popescu_Hanson_2010_MasscleanCols}, shows that 
the expected distributions of cluster colours in 
age-colour planes depend on the adopted stellar tracks: large
observed cluster samples can serve TP-AGB calibration purposes that way.
The difficulty of assigning individual clusters an age can be avoided 
by plotting $V-K$ against optical colours 
(see Fouesneau \& Lan\c{c}on, 2010, for age assignments in the 
stochastic context).
For the comparison between theoretical and observational distributions
to be meaningful, selection effects and uncertainties must be included
in the model predictions, in a way similar to what is common practice
when resolved colour-magnitude diagrams of stars in Local Group galaxies
are compared to synthetic diagrams.  The comparisons will yield constraints
not only on the TP-AGB evolutionary tracks, but also on the age and 
mass distributions of the clusters observed. With samples of 
thousands of clusters of all masses, 
I expect that first order questions such as the 
average contribution of TP-AGB stars as a function of average 
optical colours can be answered despite potential degeneracies with parameters of the 
cluster formation histories.

\section{The future : colour-magnitude diagrams, large samples,
methods that account for stochastic fluctuations}

The recent past has seen the advent of many large surveys, and this 
trend will continue. The nearest galaxies in particular will be 
studied extensively both from the ground and from space.

Resolved colour-magnitude diagrams
are expected to reduce errors in the ages assigned to individual clusters
in the nearest galaxies.  Ground-based optical colour-magnitude diagrams
for about 1500 clusters in the Magellanic Clouds have been analysed
by Glatt et al. (2010). Using overlap between their sample and others, 
they showed that ages based on colour-magnitude diagrams are 
more robust than those based on integrated colours (that do not as yet account
for the discrete nature of the stellar populations of a cluster): 
the dispersion between authors is below 0.2 in log(age) in the first case, 
of 0.3 in the second.
However, the overlap between samples is typically of 50-100 objects, which
tend to be the most luminous, least contaminated ones. For the majority 
of the clusters, the isochrone fitting techniques rely to a large extent 
on a very small number of evolved stars, whose membership is uncertain. 
Such observations already provide a means of testing age-dating techniques
based on the integrated colours of discrete synthetic clusters. The
near-IR Magellanic Cloud survey VISTA-VMC 
\citep[PI. M.-R. Cioni, see][]{Kerber_etal_2009_vista}
in particular will help. But the problems of crowding and
contamination will be difficult to solve other than statistically.

The ACS Nearby Galaxy Survey and its near-IR follow-up with HST/WFC3
(Cycle 17 proposal, PI J. Dalcanton) will 
be a mine of constraints on AGB evolution at various metallicities.
WFC3 data will also allow the definition of star cluster samples, that
will complement field star data in constraining the AGB. WFC3
has already provided new cluster samples for
the 4.5\,Mpc distant spiral M\,83 \citep{Chandar_etal_2010_M83}. The
analysis of the optical colours of these clusters with discrete models
is started (Fouesneau et al., in preparation). The natural next
step will be to use the optical vs. near-IR colour-colour distributions of the 
clusters to test AGB models. 


\bibliographystyle{asp2010}
\bibliography{ALancon_GalAGB2010}

\end{document}